\documentclass[
]{ceurart}

\usepackage{floatrow}
\usepackage{caption, subcaption, booktabs}

\newfloatcommand{capbtabbox}{table}[][]

\def\baselinestretch{0.95}
\usepackage{mdframed}
\usepackage{tikz}

\usepackage{enumitem}
\setlist{nolistsep,leftmargin=*}
\setitemize[1]{label=---}
\setenumerate[1]{label=\textcircled{\scriptsize\Alph*},
  font=\sffamily}           
            
\begin{document}

\copyrightyear{2021}
\copyrightclause{Copyright for this paper by its authors.
  Use permitted under Creative Commons License Attribution 4.0
  International (CC BY 4.0).}
\conference{In: F.B. Aydemir, C. Gralha, S.  Abualhaija, T. Breaux, M. Daneva, N. Ernst, A. Ferrari, X. Franch, S. Ghanavati, E. Groen, R. Guizzardi, J. Guo, A. Herrmann, J. Horkoff, P. Mennig, E. Paja, A. Perini, N. Seyff, A. Susi, A. Vogelsang (eds.): Joint Proceedings of REFSQ-2021 Workshops, OpenRE, Posters and Tools Track, and Doctoral Symposium, Essen, Germany, 12-04-2021}


\title{CiRA: A Tool for the Automatic Detection of Causal Relationships in Requirements Artifacts}

\author[1]{Jannik Fischbach}[
email=jannik.fischbach@qualicen.de,
orcid=0000-0002-4361-6118
]
\address[1]{Qualicen GmbH, Germany}
\address[2]{Blekinge Institute of Technology, Sweden}

\author[2]{Julian Frattini}[
email=julian.frattini@bth.se,
orcid=0000-0003-3995-6125
]
\address[3]{University of Cologne, Germany}

\author[3]{Andreas Vogelsang}[
email=vogelsang@cs.uni-koeln.de,
orcid=0000-0003-1041-0815
]

\begin{abstract}
  Requirements often specify the expected system behavior by using causal relations (e.g., If A, then B). Automatically extracting these relations supports, among others, two prominent RE use cases: automatic test case derivation and dependency detection between requirements. However, existing tools fail to extract causality from natural language with reasonable performance. In this paper, we present our tool CiRA (\textbf{C}ausality detection \textbf{i}n \textbf{R}equirements \textbf{A}rtifacts), which represents a first step towards automatic causality extraction from requirements. We evaluate CiRA on a publicly available data set of 61 acceptance criteria (causal: 32; non-causal: 29) describing the functionality of the German Corona-Warn-App. We achieve a macro $F_1$ score of 83~\%, which corroborates the feasibility of our approach.
\end{abstract}

\begin{keywords}
  Causality \sep
  Requirements Engineering \sep
  Tool Demo \sep
  Natural Language Processing
\end{keywords}

\maketitle
\vspace{-.2cm} 
\section{Introduction}
\vspace{-.2cm} 
Conditional clauses are prevalent for the specification of  system behavior, e.g., ``If the user enters an incorrect password, an error message shall be displayed'' (REQ 1). Such conditionals are a widely used linguistic pattern in both traditional requirements documents~\cite{fischbachREFSQ} as well as agile requirement artifacts, such as acceptance criteria~\cite{fischbachICST}. Semantically, conditional clauses can be understood as a causal relation between the antecedent (in case of REQ 1: \textit{incorrect password is entered}) and the consequent (\textit{error message is displayed}). Understanding and extracting such causal relations offers great potential for Requirements Engineering (RE) as it supports among others the following use cases~\cite{fischbachRE}:

\textbf{Use Case 1: Automatic Test Case Derivation From Requirements}
To derive test cases from natural language (NL) requirements, the specified system behavior and the combinatorics behind the requirement need to be understood. Specifically, we need to understand the embedded causal relation to determine the correct combination of test cases that cover all positive and negative scenarios. Currently, practitioners have to extract the causal relation manually and determine the combinations of causes and effects that need to be covered by test cases~\cite{Garousi18}. This is not only cumbersome but also becomes increasingly error-prone with growing requirements complexity~\cite{FischbachESEM}. We argue that causality extraction combined with existing automatic test case derivation methods contributes to the alignment of RE and testing (e.g., by mapping the causal relation to a Cause-Effect-Graph from which test cases can be derived automatically~\cite{fischbachICST}). 

\textbf{Use Case 2: Automatic Dependency Detection Between Requirements} 
As modern systems are becoming more and more complex, the number of requirements and their relations is constantly increasing. Practitioners fail in keeping an overview of the relationships between the requirements~\cite{fabiano18}. This may lead to undetected redundancies and inconsistencies within the requirements and consequently to faults in the system design~\cite{Vogelsang20}. We argue that an automatic causality extraction from requirements can help to compare the semantics by analyzing the different embedded causal relations. As a result, relations between requirements can be identified automatically (e.g., contradictory and redundant requirements). 

Existing approaches~\cite{Asghar16} fail to extract causality from NL with a performance that allows for use in practice. Therefore, we argue for the need of a novel method for the extraction of causality from requirements~\cite{frattini20}. We understand causality extraction as a two-step problem: We first need to detect whether requirements contain causal relations. Second, if they contain causal relations, we need to locate and extract them. In this paper, we present a demo of our tool CiRA (\textbf{C}ausality detection \textbf{i}n \textbf{R}equirements \textbf{A}rtifacts), which forms a first step towards causality extraction from NL requirements~\cite{fischbachREFSQ}. In the remainder of this paper we provide an overview\footnote{In this paper, we only provide a high level overview of the architecture of CiRA. For a detailed description of the training and tuning of our approach, please refer to our paper at the REFSQ Research Track~\cite{fischbachREFSQ}} of the functionality of CiRA (Section 2) and outline how we plan to conduct the demo at the workshop (Section 3). 
\vspace{-.1cm} 
\section{The CiRA Approach}
\vspace{-.2cm} 
CiRA has been trained to solve causality detection as a binary classification problem. Specifically, CiRA is capable of classifying single sentences or even multiple concatenated sentences written in unrestricted natural language into two categories: 1) the input contains a causal relation or 2) the input does not contain a causal relation. The classification is performed in four steps:

\begin{enumerate}
\item \textbf{Tokenization of the Text Input}
First, the text input must be decomposed into individual tokens. Since we use the Bidirectional Encoder Representations from Transformers (BERT) model~\cite{devlin19} as the foundation for CiRA, the input must be brought to a fixed length (maximum 512 tokens). For sentences that are shorter than this fixed length, padding tokens (PAD) are inserted to adjust all sentences to the same length. In addition to the PAD token, SEP tokens are inserted as special separator tokens and the CLS (classification) token is inserted as the first token in the input sequence. The CLS token represents the whole sentence (i.e., it is the pooled output of all tokens of a sentence). Our experiments~\cite{fischbachREFSQ} revealed that CiRA performs best with a fixed length of 384 tokens. 

\item \textbf{Enriching the Text Input with Syntactic Information}
In this step, we provide knowledge about the grammatical structure of the sentence to the classifier. For this purpose, we add the corresponding Dependency (DEP) tag to each token by using the spaCy NLP library. Our experiments~\cite{fischbachREFSQ} demonstrated that adding the DEP tags leads to a performance gain compared to adding Part-of-Speech (POS) tags to the text input or the usage of the vanilla BERT model.

\item \textbf{Generate Sentence Embedding}
After the pre-processing, the tokens are fed into the BERT model, which generates the corresponding embeddings. For our classification tasks, we are mainly interested in the CLS token and its embedding. Since the CLS token represents the whole sentence, the embedding created by BERT represents a sentence embedding that can be easily used for classification.

\item \textbf{Softmax Classification}
Finally, the sentence embedding is fed into a single-layer feed forward neural network that uses a softmax layer, which calculates the probability that a sentence is causal or not.
\end{enumerate}

\begin{figure}
\begin{floatrow}
\capbtabbox{%
\scriptsize
\begin{tabular}{p{5cm}c} 
\toprule
\textbf{Acceptance Criterion}                                                                         & \multicolumn{1}{l}{\textbf{Label}}  \\
The prompt no longer appears after the first time the app is used. & 1                                   \\
The terms of use can be displayed within the app.                  & 0                                   \\
The consent prompt is shown only the first time a user launches the app. & 1
\\ 
An explanation of the app’s various functions will be provided. & 0
\\
There is a “Publication information” item in the menu. & 0
\\
The IDs can be sent to the Warn server pseudonymized. & 0
\\
\bottomrule
\end{tabular}

\bigskip
\begin{tabular}{p{2cm}ccc} 
\toprule
                  & \multicolumn{1}{l}{\textbf{Precision}} & \multicolumn{1}{l}{\textbf{Recall}} & \multicolumn{1}{l}{\textbf{F1}}  \\
\textbf{Causal} \\ (Support: 32)            & 0.92                                    & 0.75                                 & 0.83                                                                      \\
\textbf{Non-causal} \\ (Support: 29)       & 0.77                                   & 0.93                                 & 0.84                                                                      \\ 
\hline
\textbf{Accuracy} \\ (Support: 61) & \multicolumn{3}{c}{0.84}                                                                                                                                   \\
\bottomrule
\end{tabular}
}{%
  \caption{Data Set (above) and Evaluation Results (below)}\label{Tabs}%
}
\ffigbox[\Xhsize]{%
\scalebox{0.4}{ 
  \includegraphics[page=1]{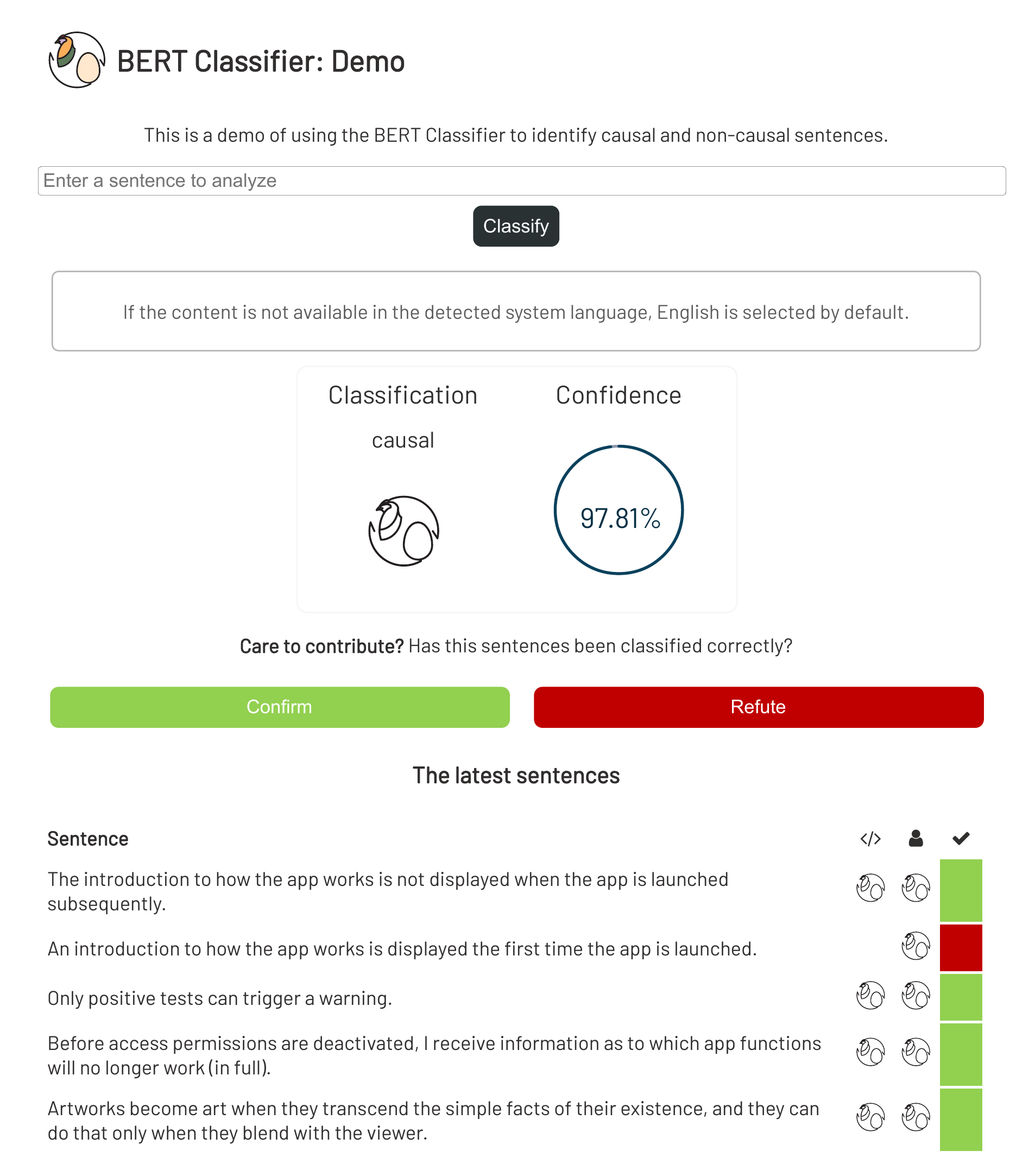}}%
}{%
  \caption{UI of CiRA}\label{UI}%
}
\end{floatrow}
\end{figure}  
  
\section{Demo Plan}
\vspace{-.2cm} 
\paragraph{Technical Setup \& User Interface}
During the workshop we will use our online demo of CiRA (\url{www.cira.bth.se\\bert}). 
The website is built as a restful node.js server utilizing the Express framework. The backend's main purpose is to execute a Python script, which acts as a wrapper around the classifier: our pre-trained binary-file classifier is loaded, the sentence classified, and the resulting classification alongside the classifier's confidence returned.
The UI provides a text input field, where an arbitrary NL sentence can be entered (see top of Fig.~\ref{UI}). On pressing the ``classify''-button, the sentence is sent to the backend where it is processed by the aforementioned, wrapped classifier. On return of the REST call, the classification and confidence of the model are rendered in the UI.
The user may confirm or correct the classifiers choice. The entered sentence and the optional user confirmation or correction is then stored in the backend, in order to (1) display the five most recently entered sentences (see bottom of Fig.~\ref{UI}), (2) provide preliminary insight into the performance of the classifier on unseen sentences, and (3) preserve sentences for future training of the classifier. Currently, we only support batch learning, but we plan to implement an online learning algorithm in future research to leverage the collected data directly for enhancing CiRA.
\vspace{-.2cm} 
\paragraph{Evaluation On Unseen Real World Data}
In the NLP4RE-workshop, we will demonstrate that CiRA is suitable for practical use. For this purpose, we evaluate CiRA on unseen real word data to simulate its application in the intended context. We use a publicly available data set of acceptance criteria\footnote{The data set can be found at \url{https://github.com/corona-warn-app/cwa-documentation/blob/master/scoping_document.md}.} provided by SAP, which describe the functionality of the German ``Corona-Warn-App''. The data set consists of 32 user stories containing a total of 61 acceptance criteria. In order to measure the performance of CiRA, we manually annotated all acceptance criteria and classified them into two categories: causal (Label: 1) and non-causal (Label: 0). We then classified each acceptance criterion using CiRA and compared the results with ground truth (see Tab.~\ref{Tabs}). During the workshop, we will present both true and false predictions and discuss the performance, challenges, and possibilities of CiRA with the other participants.
\vspace{-.2cm} 

\bibliography{sample-ceur}

\end{document}